\documentclass[doublecol,a4paper,showpacs]{epl2}
\usepackage{graphicx}
\usepackage[all]{xy}
\usepackage{amsmath}
\usepackage{amssymb}
\usepackage{tensor}
\newcommand{\be}{\begin{equation}}
\newcommand{\ee}{\end{equation}}
\newcommand{\ben}{\begin{eqnarray}}
\newcommand{\een}{\end{eqnarray}}
\newcommand{\bes}{\begin{subequations}}
\newcommand{\ees}{\end{subequations}}
\def\bal#1\eal{\begin{align}#1\end{align}}

\newcommand{\bb}{\bibitem}

\begin{document}

\title{A novel connection between scalar field theories and quantum mechanics}

\author{D. Bazeia\thanks{Corresponding author. \email{bazeia@fisica.ufpb.br}} \and L. Losano}
\shortauthor{D.Bazeia \etal}
\institute{Departamento de F\'\i sica, Universidade Federal da Para\'\i ba, 58051-970 Jo\~ao Pessoa, PB, Brazil}
\date{\today}

\abstract{This work deals with scalar field theories and supersymmetric quantum mechanics. The investigation is inspired by a recent result, which shows how to use the reconstruction mechanism to describe two distinct field theories from the very same quantum mechanics potential, and by an older work, which describes the deformation procedure that offers an interesting way to generate and solve new scalar field theories, starting from a given model of current interest. We use the procedure to unveil a new route, from which one can describe families of scalar field theories that engender the very same quantum mechanics potential. The approach can be applied algorithmically, and implemented to generate models that give rise to distinct quantum mechanics systems as well.}


\maketitle

The study of localized structures in models described by real scalar fields in $(1,1)$ spacetime dimensions has attracted the attention of several researches along the last decades. In the book \cite{b1}, and in the specific works \cite{p00,blm,p0,p1,p10,p11,p20,p2,p3,p4,px,p5,p6,p7} and in references therein one can find a diversity of investigations related to the presence of kinks and/or domain walls of current interest to high energy physics and condensed matter.  

In high energy physics, the study of kinks and its linear stability leads us to distinct but related issues, the supersymmetric quantum mechanics \cite{susyqm1,susyqm2,susyqm3} and the reconstruction of field theory models from supersymmetric quantum mechanics; see,
e.g., Refs.~\cite{susyqm1,susyqm2,susyqm3,qm1,qm2} and references therein. In this context, in the recent works \cite{qm3,qm4} some of us reexamined the issue to show that, in the case of the reconstruction of scalar field theory from quantum mechanics, one can obtain two field theories from the very same quantum potential. This result motivated us to further explore the subject, and here we describe another possibility, which paves a distinct route of investigation of field theory models that support the very same stability potential and so the same quantum system.

\begin{figure}[h]
\begin{center}
\includegraphics[{height=2.6cm,width=6cm}]{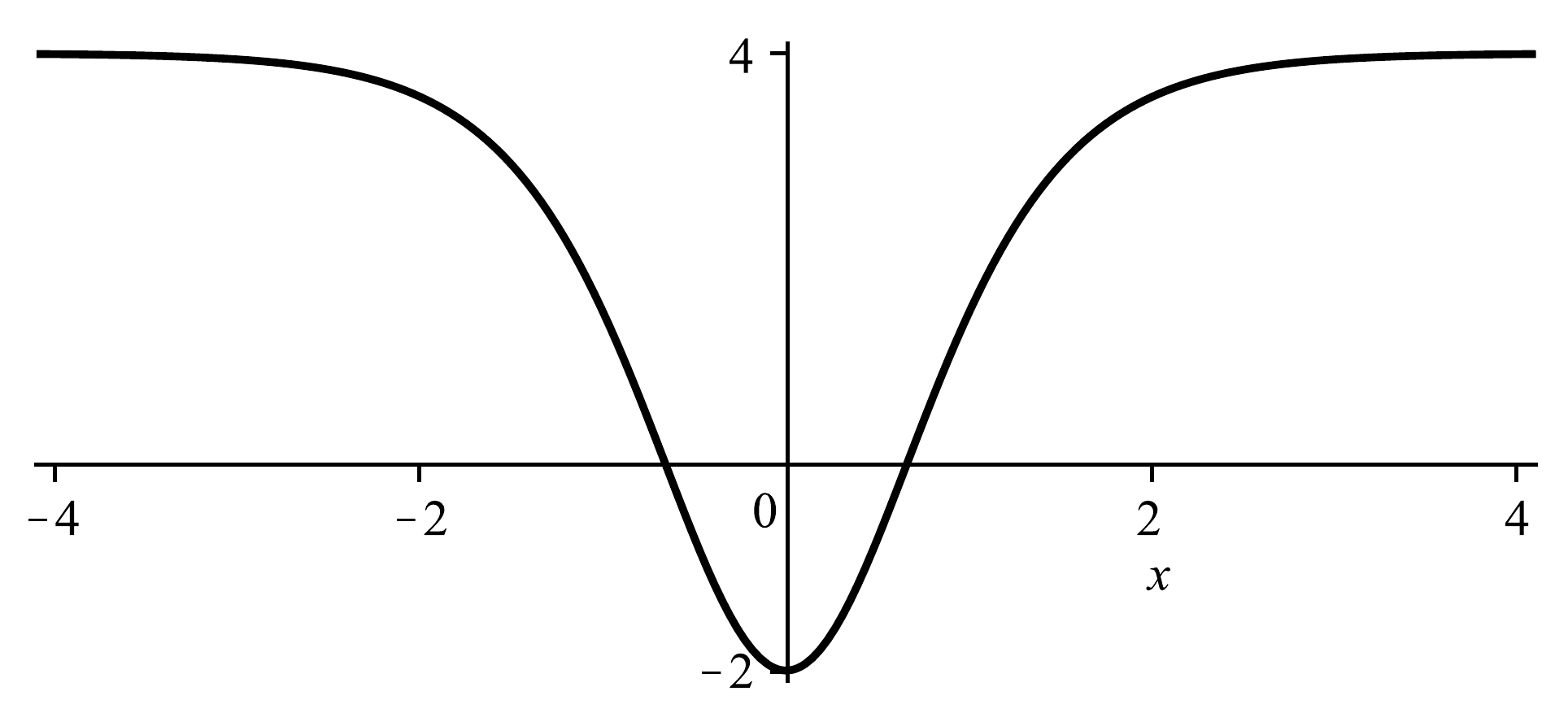}
\end{center}
\caption{The quantum mechanical potential \eqref{pt}.}\label{fig1}
\end{figure}

Before developing the procedure, let us briefly review the main result of Ref.~\cite{qm4}. Taking natural units and dimensionless fields and space and time coordinates, and using the reconstruction methodology suggested in Ref.~\cite{qm2}, one showed how to obtain the two potentials 
\be\label{V1}
V_0(\phi)=\frac12 (1-\phi^2)^2,
\ee
and
\be\label{V2}
V_1(\phi)=\frac12 \phi^2 (2-|\phi|)^2,
\ee
from the very same quantum mechanics, described by the modified P\"oschl-Teller potential \cite{PT}, written in the form
\be\label{pt}
u(x)=4-6\,{\rm sech}^2(x),
\ee
which is known to support two bound states, the zero mode with zero energy and another bound state, the vibrational state with positive energy. The modified P\"oschl-Teller potential is depicted in Fig.~\ref{fig1} and is symmetric around the origin, that is, $u(x)=u(-x)$.

\begin{figure}[t]
\begin{center}
\includegraphics[{height=3cm,width=6cm}]{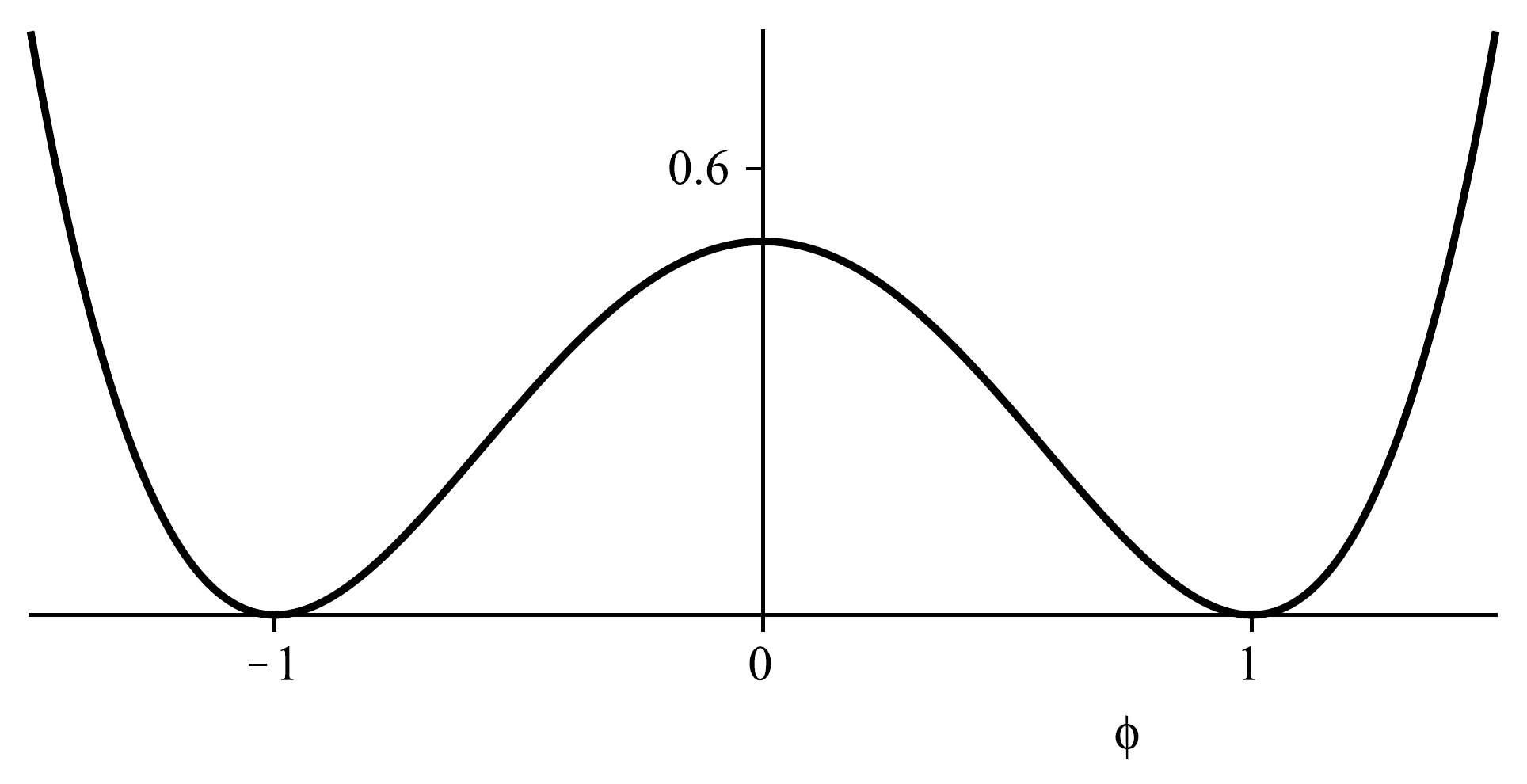}
\includegraphics[{height=3cm,width=6cm}]{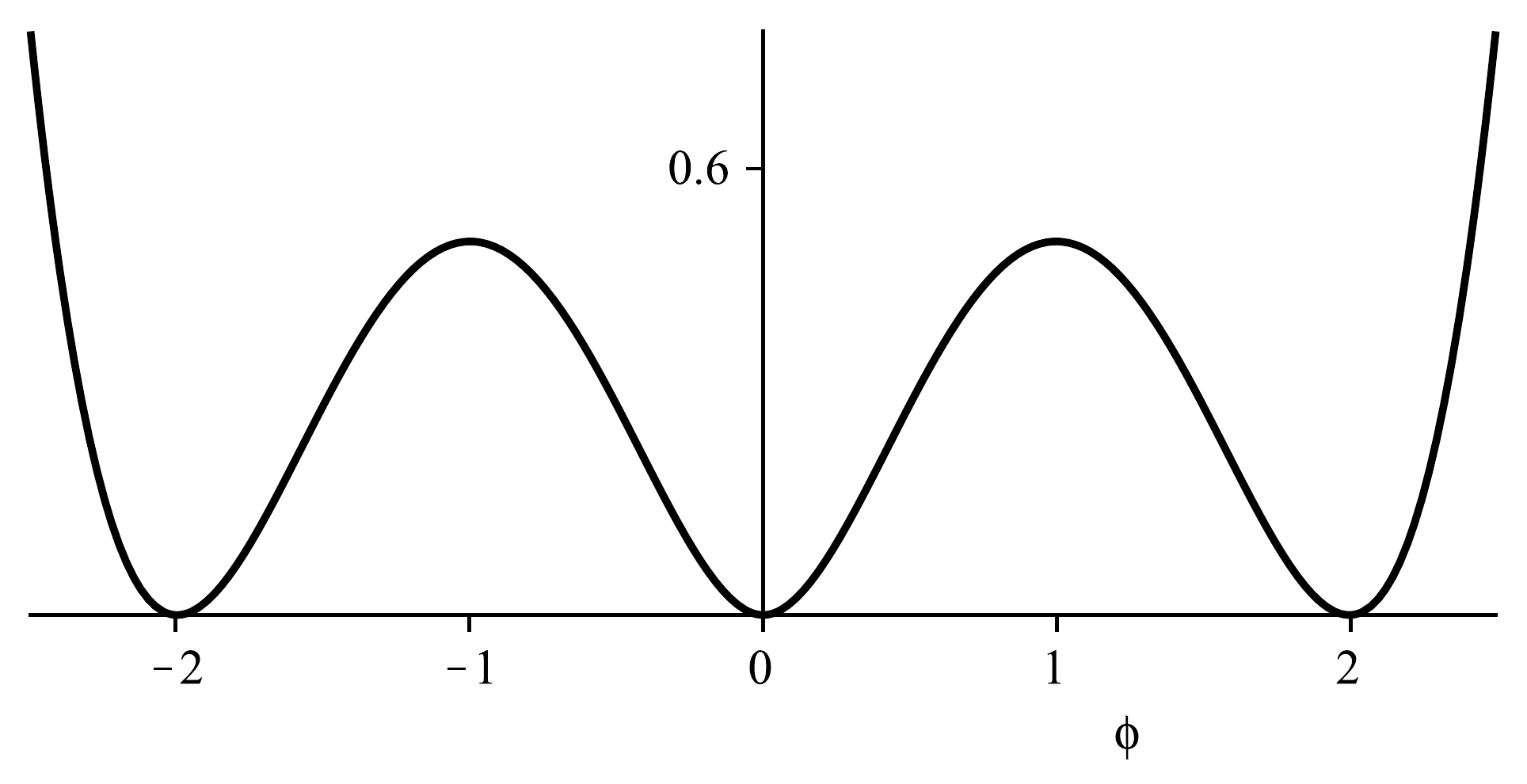}
\end{center}
\caption{The two potentials \eqref{V1} (top) and \eqref{V2} (bottom) and their topological sectors $(-1,1)$ and $(-2,0)$ and $(0,2)$, respectively.}\label{fig2}
\end{figure}

The potential \eqref{V1} is the prototype of the Higgs field potential, and engenders a single topological sector, identified by the two minima
$\phi=\pm1$ in the form $(-1,1)$. The potential \eqref{V2} was first considered in Ref.~\cite{bil} and engenders two topological sectors, identified by the three minima $\phi=0,\pm2$. They are the left $(-2,0)$ and right $(0,2)$ sectors, which are similar, symmetric around their corresponding maxima. The two models are very different from each other; while the first model may be used to describe a second-order phase transition, the second potential supports a minimum at $\phi=0$ which is symmetric, so the model may describe coexistence of symmetric and asymmetric phases, which is typical of a first-order phase transition. For clarity, we depict the potentials \eqref{V1} and \eqref{V2} in Fig.~\ref{fig2}. 

The two models can be reconstructed from the very same quantum mechanics, with the potential \eqref{pt}; see Ref.~\cite{qm4}. Additionally, we can also check that kinklike solutions that appear from $V_0(\phi)$ and $V_1(\phi)$ are
\be  
\phi_0(x)=\tanh(x),
\ee
and 
\be  
\phi_1(x)=1+\tanh(x),
\ee 
and generate the very same stability potential, which coincides with \eqref{pt}, again leading to the very same quantum system.

\begin{figure}[t]
\begin{center}
\includegraphics[{height=3cm,width=6cm}]{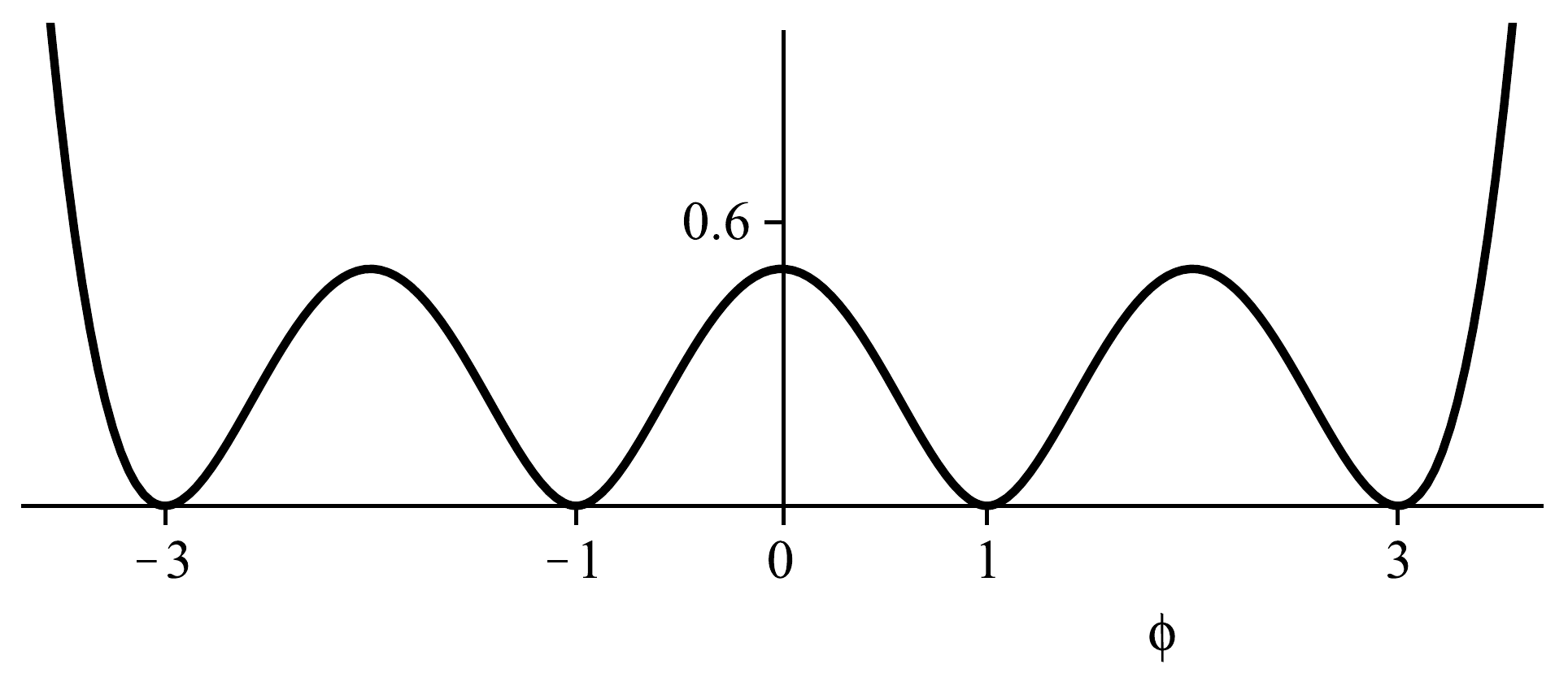}
\includegraphics[{height=3cm,width=6cm}]{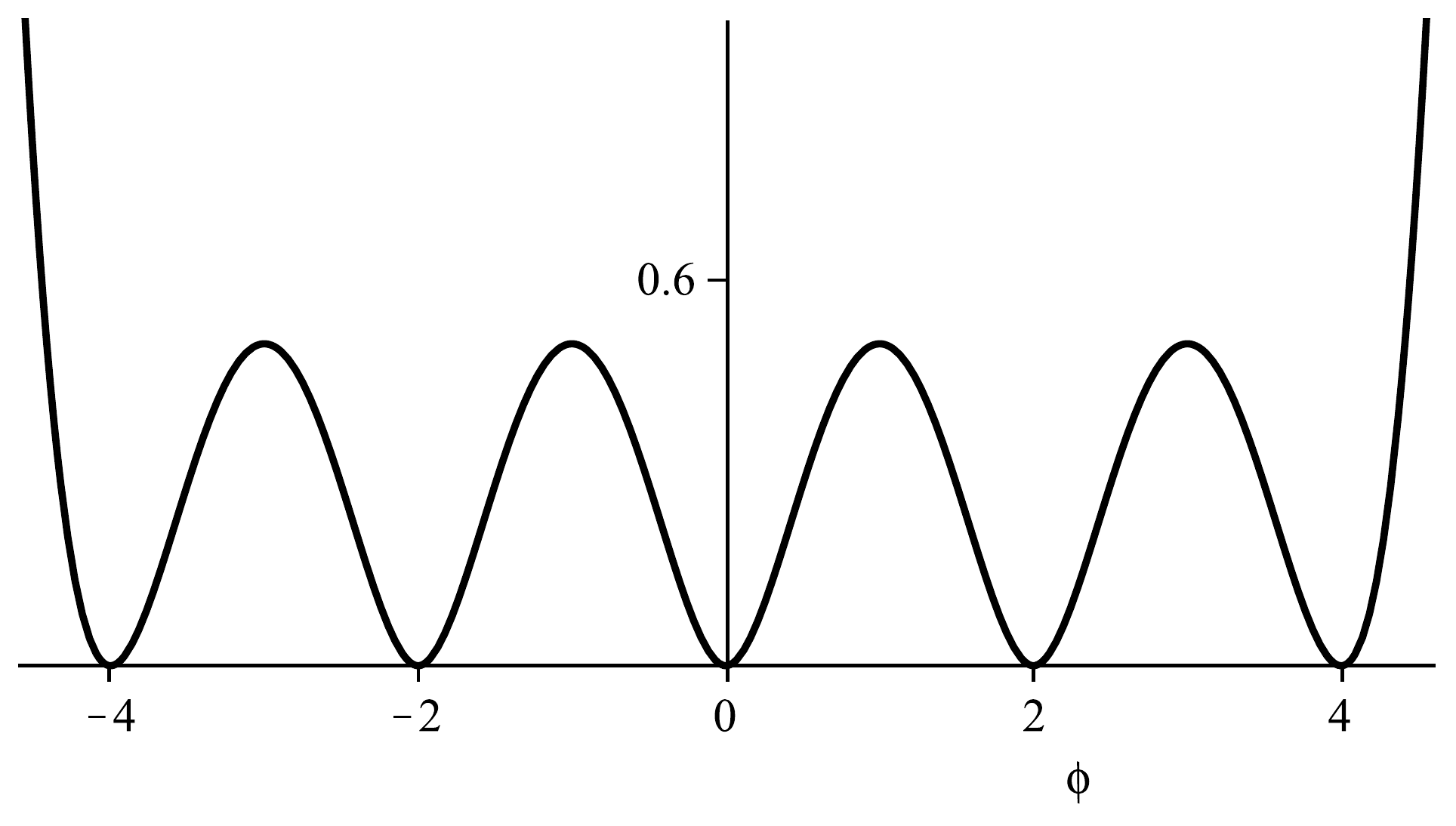}
\end{center}
\caption{The potential \eqref{pot1b}, depicted for $b=1$ and for $b=2$, in the top and bottom panels, respectively.}\label{fig3}
\end{figure}
Although the reconstruction procedure \cite{qm1,qm2,qm4} may respond correctly under appropriate circumstances, the aim of the current study is to offer an alternative route to investigate the subject. We implement this goal using the deformation procedure introduced in Ref.~\cite{blm}. This procedure has been used in several different contexts, as the ones recently developed in Refs.~\cite{D1,D2,D3,D4,D5,D6,D7}, to help describe issues related to braneworlds, inflation under the slow-roll regime and self-duality, for instance. The method starts with a given Lagrangian density, say
\be 
{\cal L}=\frac12\partial_\mu\phi\partial^\mu\phi-V(\phi).
\ee
We then choose an appropriate deformation function $f(\phi)$ and define the deformed potential
\be 
{\widetilde V}(\phi)=\frac{V(\phi\to f(\phi))}{f^{\prime\, 2}},
\ee
where $\phi\to f(\phi)$ means that one should change $\phi$ to $f(\phi)$ in the potential $V(\phi)$, and $f^\prime$ stands for the derivative of
$f$ with respect to its argument, that is, $f^{\prime}={df}/{d\phi}$. This leads us to the deformed model
\be  
{\widetilde{\cal L}}=\frac12\partial_\mu\phi\partial^\mu\phi-{\widetilde V}(\phi).
\ee

To use the deformation procedure in the current work, one first notices that the above model \eqref{V2} can be obtained from the model \eqref{V1} with the use of the function $f(\phi)=1- |\phi|$. In fact, we can be more general and use the deformation function 
\be 
f_a(\phi)=a-|\phi|,
\ee
with $a$ a non negative real number, to get from the potential \eqref{V1} to the potential
\be\label{potdef}
V_a(\phi)=\frac12(1-a^2+2a|\phi|-\phi^2)^2.
\ee
This expression shows that $a=0$ leads us back to the potential \eqref{V1}, and for $a=1$ one gets the potential \eqref{V2}. But it tells us more: the two values $a=0$ and $a=1$ coincide with the two inequivalent values of the field that are extrema of the potential \eqref{V1}, the points $\phi=0$ and $\phi=1$. The other values of $a$ are in the intervals $a\in(0,1)$ and $a\in(1,\infty)$, but they lead to potentials with discontinuity in the derivative. In this work we avoid the appearance of discontinuities and then choose $a=0$ and $a=1$, which gives rise to the two models depicted in Fig.~\ref{fig2}.

The above deformation procedure tells us much more: we can take the potential $V_1(\phi)$ and deform it again, using $f_b(\phi)=b-|\phi|$, with $b$ real and nonnegative. The procedure leads us to the new potentials 
\be\label{pot1b}
V_{1,b}(\phi)= \frac12(b-|\phi|)^2(2-|b-|\phi||)^2)^2.
\ee
We see in the bottom panel in Fig.~\ref{fig2} that $\phi=0$, $\phi=1$ and $\phi=2$ are inequivalent extrema of $V_1(\phi)$, so we can use $b=0$, $b=1$ and $b=2$, with the two last cases giving rise to two new potentials, which we display in Fig.~\ref{fig3}. 

The two potentials $V_0, V_1$ that appear from \eqref{potdef} and the two other $V_{1,1}, V_{1,2}$ that appear from \eqref{pot1b} are distinct potentials that have several topological sectors, each one of them giving rise to the very same quantum system. Moreover, it is possible to deform the two new potentials $V_{1,1}$ and $V_{1,2}$ once again, to give rise to new potentials. We then identify an algorithmic procedure, which can be used to generate two families of field theory models, one with an odd number of topological sector, and the other with an even number of sectors, all of them generating the very same quantum system.

\begin{figure}[t]
\begin{center}
\includegraphics[{height=4cm,width=6cm}]{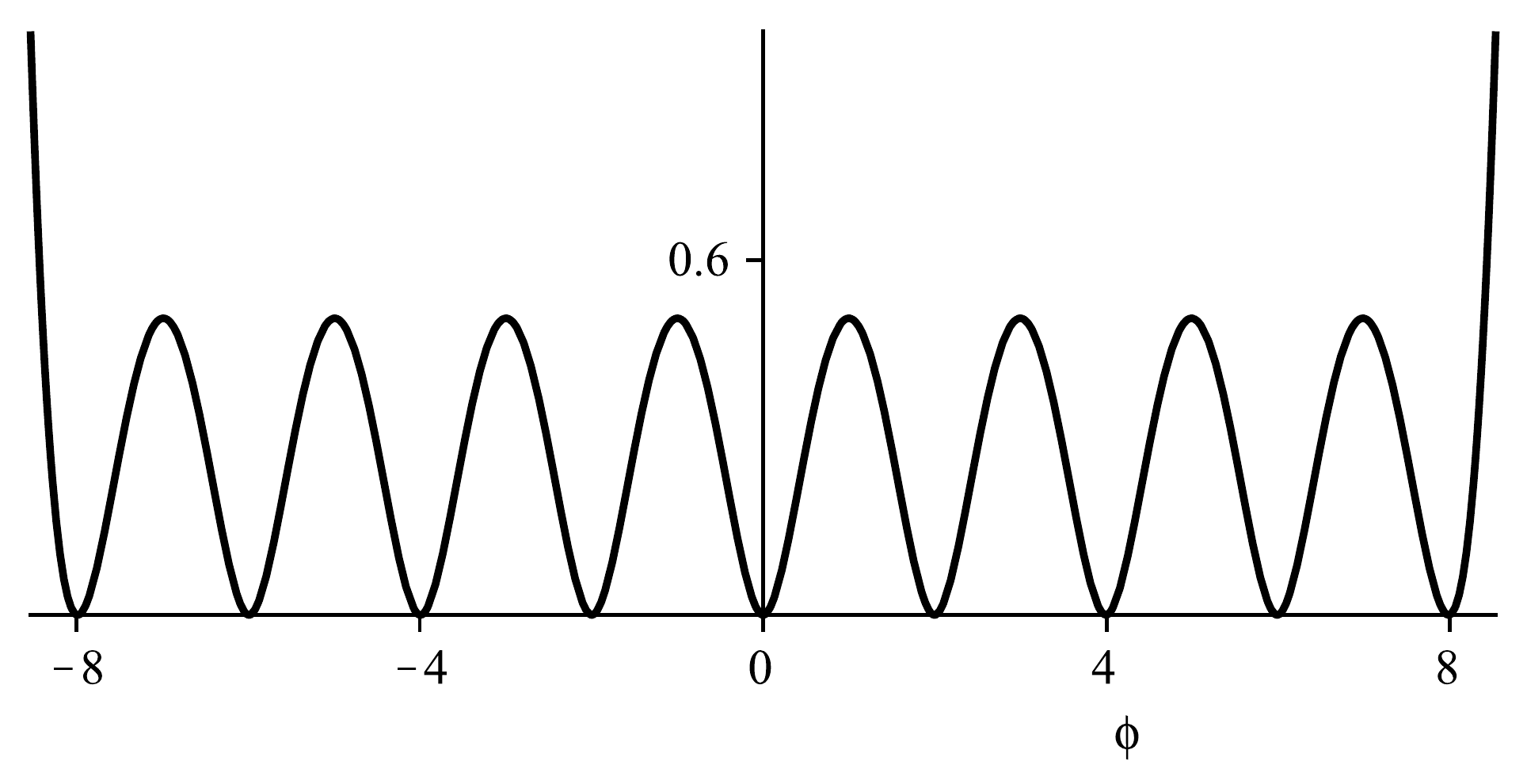}
\end{center}
\caption{The potential that appears from $V_0(\phi)$, after using the deformation procedure with $f(\phi)=1-|\phi|$ again and again, seven times.}\label{fig4}
\end{figure}

\begin{figure}[t]
\begin{center}
\includegraphics[{height=3cm,width=6cm}]{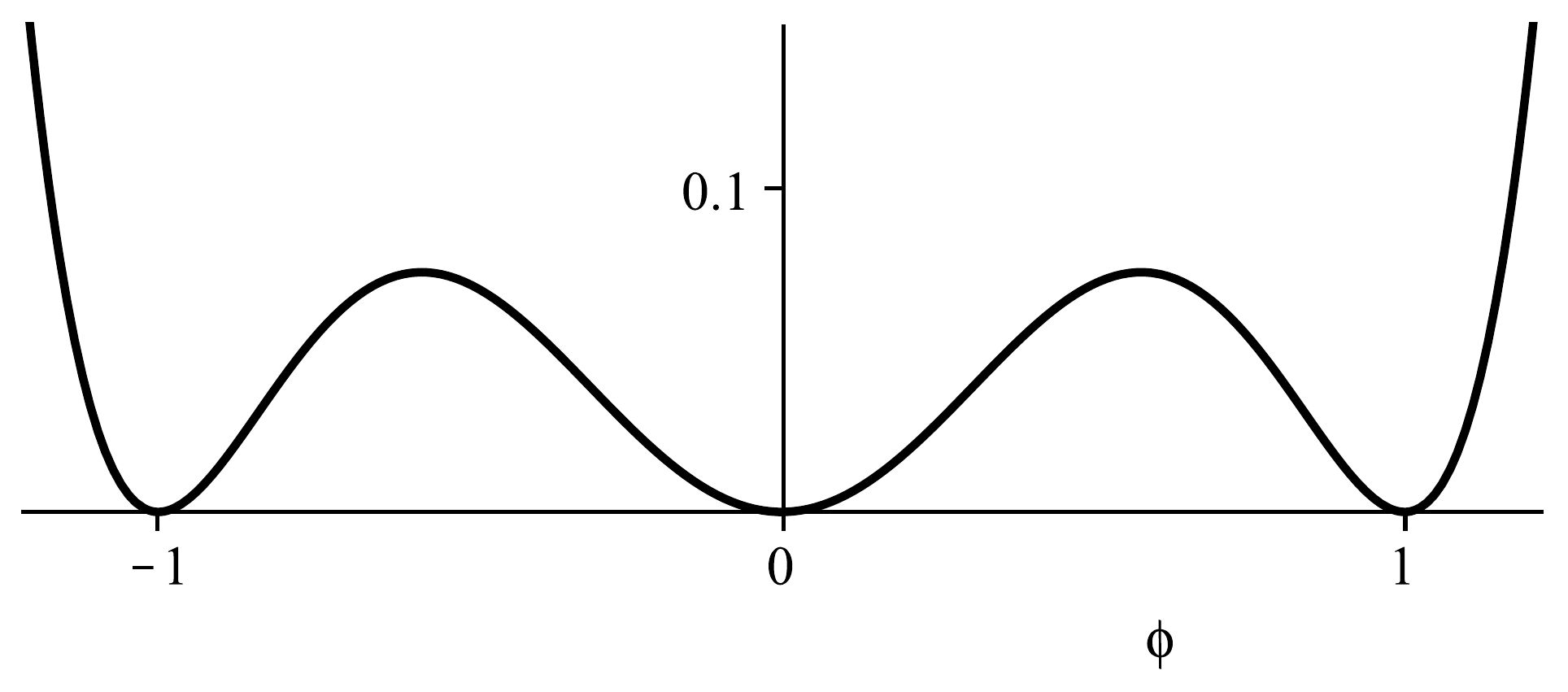}
\end{center}
\caption{The $\phi^6$ potential \eqref{V6} as a function of $\phi$ showing the two (left and right) topological sectors.}\label{fig5}
\end{figure}

\begin{figure}[t]
\begin{center}
\includegraphics[{height=3cm,width=6cm}]{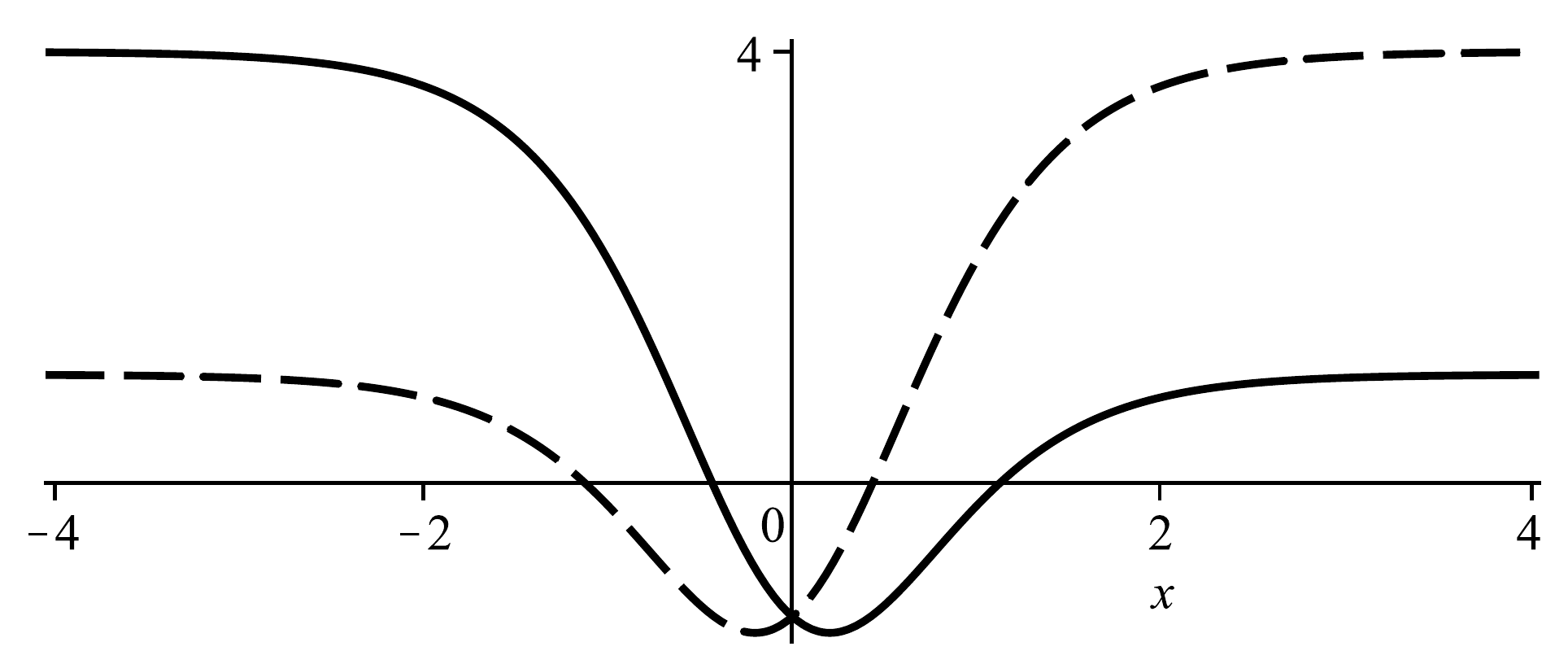}
\end{center}
\caption{The two potentials $u_l(x)$ and $u_r(x)$ that appear in \eqref{ul} and \eqref{ur}, depicted with solid and dashed lines, respectively.}\label{fig6}
\end{figure}

We illustrate the above results once again in Fig.~\ref{fig4}, displaying the potential that results from the deformation of the $V_0(\phi)$, after using the deformation function $f(\phi)=1-|\phi|$ again and again, seven times. Each topological sector gives rise to the very same quantum problem.

Another difficulty with the reconstruction procedure is that if one starts with a reflectionless potential, it can not be used to reconstruct the
$\phi^6$ model, for instance, because it engenders two asymmetric sectors which gives rise to asymmetric stability potentials and then to quantum problems that are not reflectionless. For this reason, let us now study the $\phi^6$ model from the above deformation perspective. The potential can be written as
\be\label{V6}
U_0(\phi)=\frac12\phi^2(1-\phi^2)^2\,.
\ee
It has two topological sectors, the left $(-1,0)$ and the right $(0,1)$ sectors, as illustrated in Fig.~\eqref{fig5}. But now the topological sectors are asymmetric around the maxima at
$\phi=\pm\sqrt{3}/3$, and this will generate new possibilities. The kinklike solutions are given by
\be\label{sol51}
\phi_{l}(x)=-\frac12\sqrt{2\left(1-\tanh(x)\right)}\,,
\ee
\be\label{sol52}
\phi_{r}(x)=\frac12\sqrt{2\left(1+\tanh(x)\right)}\,,
\ee
in the left and right sectors, respectively.

In this case, there are two stability potentials which are given by
\ben
{u}_l(x)&=&\frac52-\frac32\tanh(x)-\frac{15}{4}\,{\rm sech}^2(x)\,,\label{ul}\\
{u_r}(x)&=&\frac52+\frac32\tanh(x)-\frac{15}{4}\,{\rm sech}^2(x)\,,\label{ur}
\een
and are depicted in Fig.~\ref{fig6}. As it is well-known, they support the zero mode and no other bound state. Also, they are connected by  $u_r(x)=u_l(-x)$, so they represent two equivalent problems.

\begin{figure}[t]
\begin{center}
\includegraphics[{height=3cm,width=6cm}]{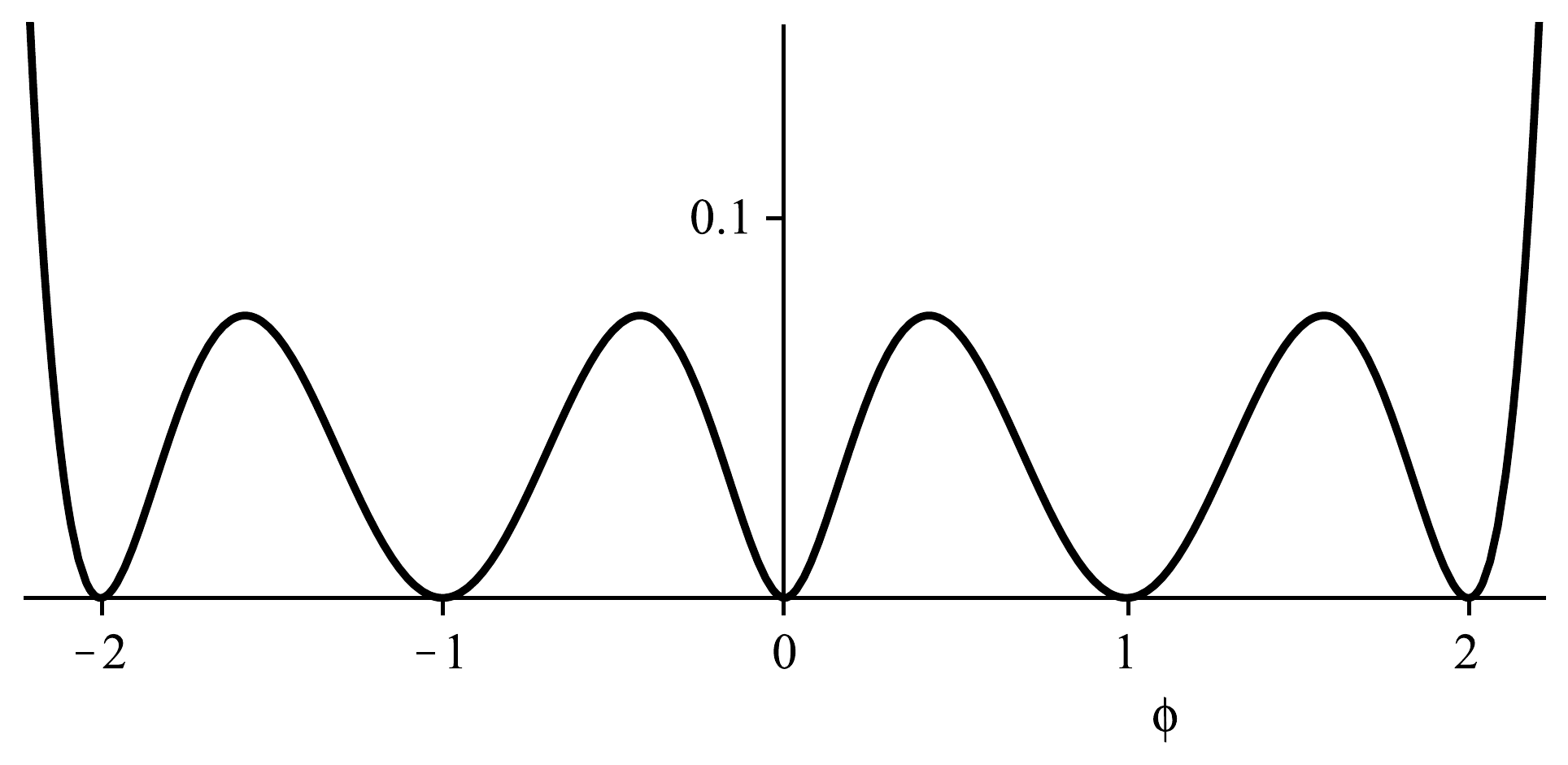}
\end{center}
\caption{The potential \eqref{pot6d}.}\label{fig7}
\end{figure}

We consider the potential \eqref{V6} and the deformation function $f_c(\phi)=c-|\phi|$ in order to obtain 
\be\label{v51}
U_c(\phi)=\frac12(c-|\phi|)^2\left(1-c^2+2c|\phi|-\phi^2\right)^2\,.
\ee
The potential \eqref{V6} has extrema at $0$, $\pm\sqrt{3}/3$, and $\pm1$. If one uses $c=0$, the procedure leads us back to the potential \eqref{V6}, so one is left with the cases $c=1$ and $c=\sqrt{3}/3$. We take $c=1$ to get
\be\label{pot6d}
U_1(\phi)=\frac12\phi^2(1-|\phi|)^2(2-|\phi|)^2\,,
\ee
which has four topological sectors, as illustrated in Fig.~\eqref{fig7}. The kinklike solutions are
\be\label{sol61}
\phi_{l1}(x)=-1-\frac12\sqrt{2\left(1-\tanh(x)\right)}\,,
\ee
\be\label{sol62}
\phi_{l2}(x)=-1+\frac12\sqrt{2\left(1+\tanh(x)\right)}\,,
\ee
\be\label{sol63}
\phi_{r1}(x)=1-\frac12\sqrt{2\left(1-\tanh(x)\right)}\,,
\ee
and
\be\label{sol64}
\phi_{r2}(x)=1+\frac12\sqrt{2\left(1+\tanh(x)\right)}\,.
\ee
They appear in the topological sectors labelled $l1, l2, r1, r2$, identified by the several minima $(-2,-1), (-1,0), (0,1), (1,2)$, respectively.

The two topological sectors $l1$ and $r1$ are related to the solutions \eqref{sol61} and \eqref{sol63}, respectively, and give rise to the potential \eqref{ul}. The other two topological sectors $l2$ and $r2$, related to the solutions \eqref{sol62} and \eqref{sol64}, give rise to the potential \eqref{ur}. So we can say that the potentials $U_0(\phi)$ and $U_1(\phi)$ give rise to the very same quantum mechanics.

\begin{figure}[ht]
\begin{center}
\includegraphics[{height=3cm,width=6cm}]{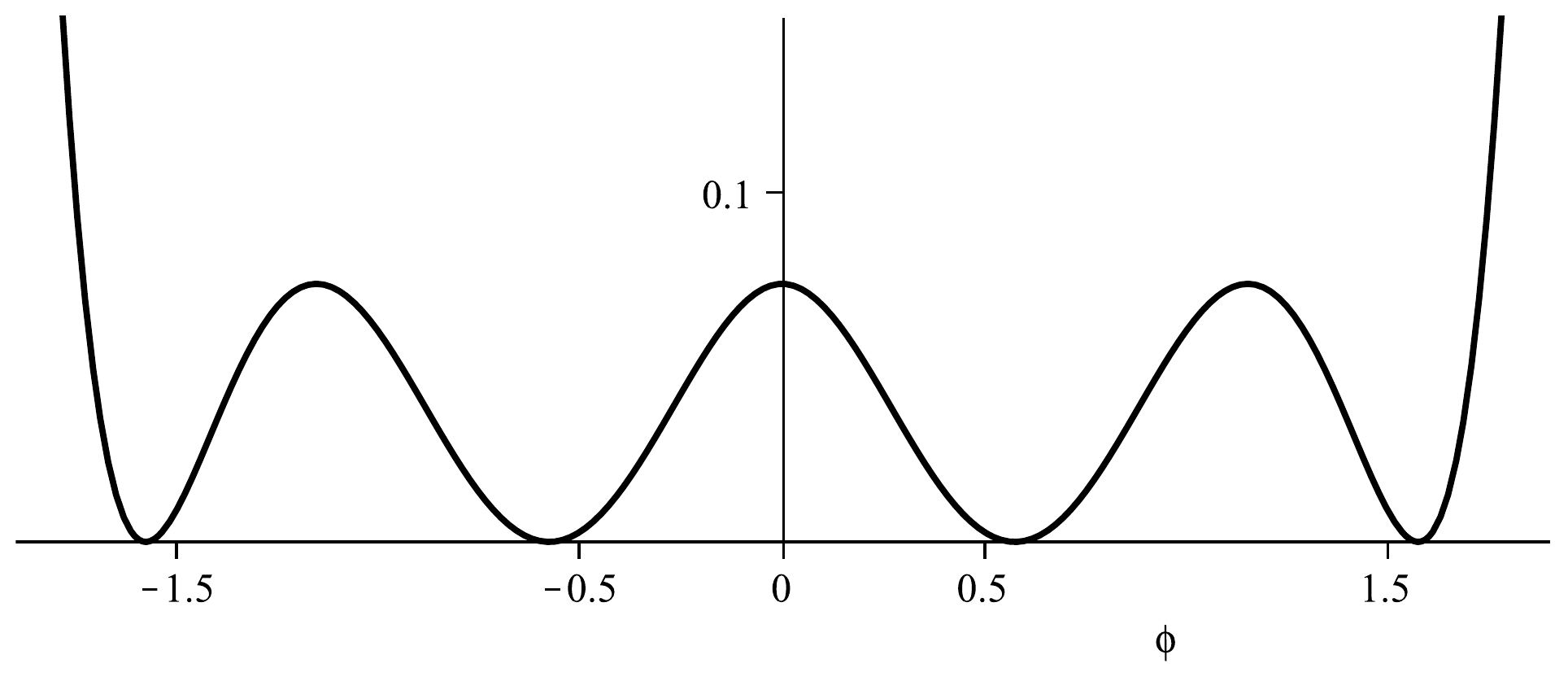}
\end{center}
\caption{The potential \eqref{v51} depicted for $c=\sqrt{3}/3$.}\label{fig8}
\end{figure}

However, the case with $c=\sqrt{3}/3$ is different. The potential \eqref{v51} for $c=\sqrt{3}/3$ is illustrated in Fig.~\eqref{fig8}. In this case there are three topological sectors, the left, the right and the central sectors.

\begin{figure}[t]
\begin{center}
\includegraphics[{height=3cm,width=7cm}]{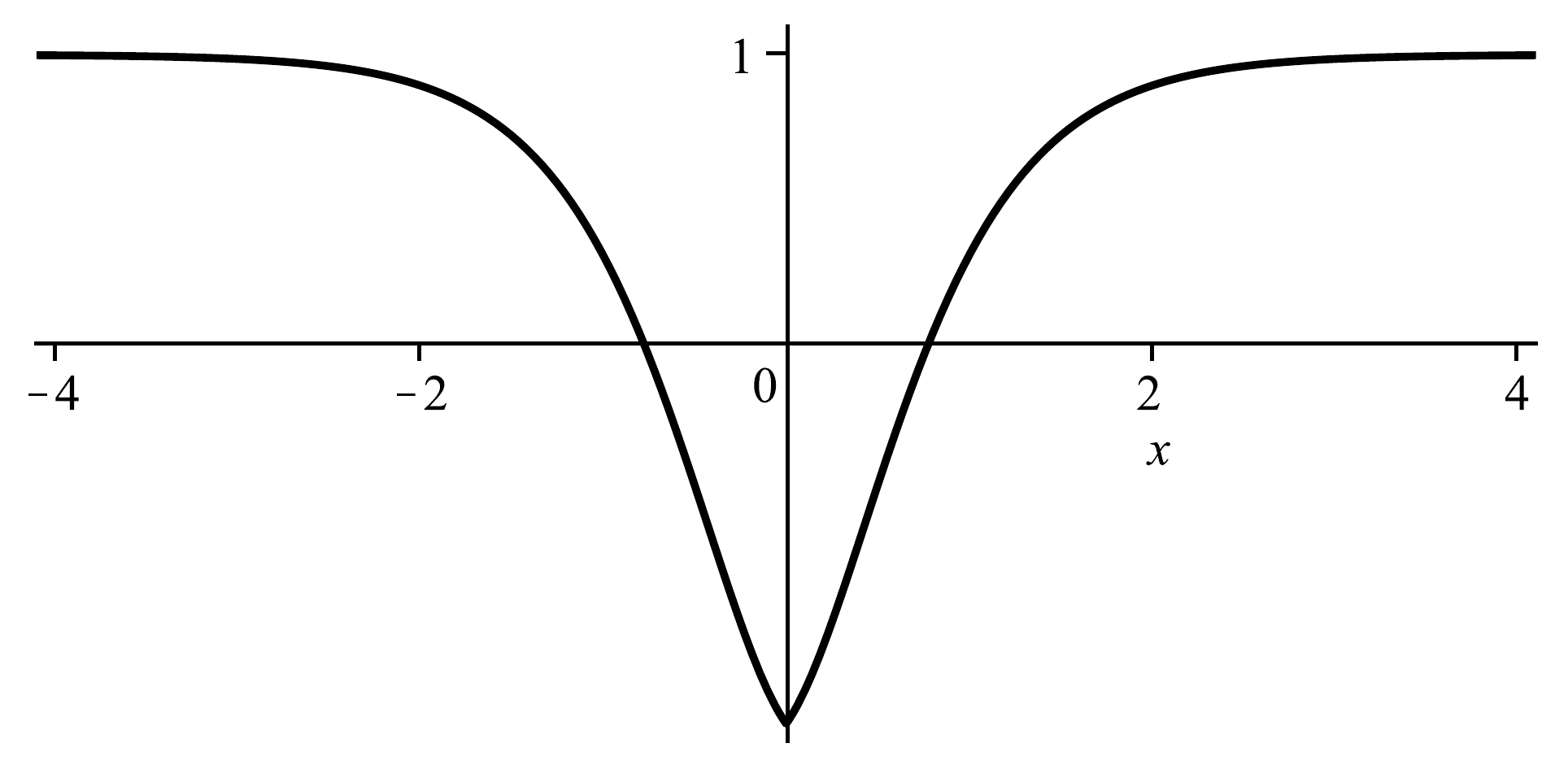}
\end{center}
\caption{The potential \eqref{qp6} that appear connected to the central sector of the model depicted in Fig.~\ref{fig8}.}\label{fig9}
\end{figure}

In the left and right sectors, the kinklike solutions are
\be\label{sol71}
\phi_l(x)=-\frac{\sqrt{3}}{3}-\frac12\sqrt{2\left(1-\tanh(x)\right)}\,,
\ee
\be\label{sol72}
\phi_r(x)=\frac{\sqrt{3}}{3}+\frac12\sqrt{2\left(1+\tanh(x)\right)}\,,
\ee
and they give rise to the two potentials \eqref{ul} and \eqref{ur}, respectively.
For the central topological sector the kinklike solution can be written as 
\begin{equation}\label{sol7c}
{\phi}(x)=
\left\{
\begin{array}{c}
-\frac{\sqrt{3}}{3}+\frac12\sqrt{2\left(1+\tanh(x-x_0)\right)},\;\;\;\;\;x\leq 0\,,\\
\,\,\\
\,\,\,\,\frac{\sqrt{3}}{3}-\frac12\sqrt{2\left(1-\tanh(x+x_0)\right)},\;\;\;\;\;x\geq 0\,,
\end{array}
\right.
\end{equation}
where $x_0={\rm arctanh}(1/3)$. This solution gives rise to another quantum mechanical potential, which is given by
\begin{equation}\label{qp6}
u(x)=
\left\{
\begin{array}{c}
\frac52+\frac32\tanh(x-x_0)-\frac{15}{4}\,{\rm sech}^2(x-x_0),\;x\leq 0\,,
\\\,\,\\
\frac52-\frac32\tanh(x+x_0)-\frac{15}{4}\,{\rm sech}^2(x+x_0),\;x\geq 0\,.
\end{array}
\right.
\end{equation}
This potential is illustrated in Fig.~\eqref{fig9}, and it supports the zero mode and no other bound state. It represents a new potential, symmetric, different from the previous ones.

We can go on, as we did in the case of the $\phi^4$ model \eqref{V1}, to construct two distinct families of models, one with an even number of topological sectors, having equivalent potentials, as it happened with the $U_1$ model \eqref{pot6d}, and the other with an odd number of sectors, with the two external sectors as the left and right sectors of the $\phi^6$ model, and one, three, five, and so on, internal sectors, similar to the central sector of the model displayed in Fig.~\ref{fig8}, having the very same potential \eqref{qp6} which is depicted in Fig.~\ref{fig9}. 

In summary, the above results show that if one starts with a scalar field theory that engenders a topological sector that is symmetric, it will give rise to a reflectionless potential. In this case, the deformation procedure proposed above can be used to generate families of models having the very same stability potential and so the same quantum mechanics. However, if one starts with a field theory that engenders two asymmetric topological sectors, we can generate two distinct families, one with an even number of topological sectors, having the very same quantum mechanics, and the other with an odd number of topological sectors. In the case of the odd number of sectors, one finds a new topological sector, the central sector which is different from the other asymmetric sectors and gives rise to a new quantum system that is reflectionless. 

The approach developed above can be used with other polynomial potential, to generate new families of potentials and their corresponding quantum systems. In this sense, the works in Refs.~\cite{new1,new2} introduce several models of real scalar fields with a diversity of topological sectors, and we are now examining how to deform them to obtain new families of models with similar quantum mechanics \cite{longer}. We have done the same with the sine-Gordon model, but since it is periodic and supports an infinite number of equivalent topological sectors, we have gotten nothing new, because adding a finite number of topological sectors to it does not change the model itself. We speculate that such property may be related to integrability of the sine-Gordon model, but this requires further examination. The procedure suggested in this work may represent a new avenue for the study of scalar field theories and supersymmetric quantum mechanics, and we will further report on this in the near future. 

The authors would like to thank Conselho Nacional de Desenvolvimento Cient\'\i fico e Tecnol\'ogico (CNPq) for partial financial support. 


\end{document}